\documentclass[revtex4]{emulateapj}

\usepackage{amssymb}
\usepackage{amsmath}
\usepackage[]{graphicx}
\usepackage{enumerate}
\usepackage{subeqnarray}
\usepackage{cases}
\usepackage{mathrsfs,amssymb}
\usepackage[usenames]{color}

\tightenlines

\begin{document}

\title{Gas and dust temperature in pre-stellar cores revisited: \\New limits on cosmic-ray ionization rate}

\author{Alexei V. Ivlev$^1$, Kedron Silsbee$^1$, Olli Sipil\"a$^1$, Paola Caselli$^1$}
\email[e-mail:~]{ivlev@mpe.mpg.de} \email[e-mail:~]{ksilsbee@mpe.mpg.de} \affiliation{$^1$Max-Planck-Institut f\"ur
Extraterrestrische Physik, 85748 Garching, Germany }

\begin{abstract}
We develop a self-consistent model for the equilibrium gas temperature and size-dependent dust temperature in cold, dense
pre-stellar cores, assuming an arbitrary power-law size distribution of dust grains. Compact analytical expressions
applicable to a broad range of physical parameters are derived and compared with predictions of the commonly used standard
model. It is suggested that combining the theoretical results with observations should allow us to constrain the degree of
dust evolution and the cosmic-ray ionization rate in dense cores, and to help in discriminating between different regimes of
cosmic-ray transport in molecular clouds. In particular, assuming a canonical MRN distribution of grain sizes, our theory
demonstrates that the gas temperature measurements in the pre-stellar core L1544 are consistent with an ionization rate as
high as $\sim 10^{-16}$~s$^{-1}$, an order of magnitude higher than previously thought.
\end{abstract}


\maketitle

\section{Introduction}
\label{intro}

Pre-stellar cores set the initial conditions for the process of star formation \citep[see, e.g.,][]{Shu1987,Bergin2007}, and
therefore determine the properties of the future stars and stellar systems which will form in their centers. Unveiling the
physical and chemical structure of the cores puts stringent constraints on dynamical/chemical models
\citep[e.g.,][]{Keto2014,Keto2015,Vasyunin2017, Sipila2018, Caselli2019}. Specifically, measuring volume density and
gas/dust temperature profiles is crucial to provide information about the heating by the interstellar radiation field (ISRF)
\citep[e.g.,][]{Zucconi2001,Launhardt2013,Steinacker2016, Harju2017, Hocuk2017} and cosmic rays (CRs)
\citep[e.g.,][]{Goldsmith2001,Keto2010}, as well as about dust evolution \citep[][]{Sadavoy2016,Chacon2019}. These processes
critically affect the dynamical and chemical evolution of pre-stellar cores, as they regulate the ionisation fraction
\citep[e.g.,][]{McKee1989} and surface chemistry/freeze-out rates \citep[e.g.,][]{Zhao2018,Shingledecker2018}.

The physical characteristics of pre-stellar cores exhibit significant variations. Typical values of the gas density $n_{\rm
g}$ in the center of pre-stellar cores exceed $\sim10^5$~cm$^{-3}$ \citep[e.g.,][]{Keto2008,Keto2015}. The gas cooling in
this case is determined by collisions with dust grains, while the direct cooling by CO line emission (dominating at lower
densities) becomes inefficient \citep[][]{Goldsmith2001,Galli2002}. Densities of up to $\sim10^7$~cm$^{-3}$ have been
reported \citep[][]{Keto2010,Caselli2019}, while the actual peak density in contracting cores is expected to reach much
higher values before a protostar forms. The magnitude of the CR ionization rate $\zeta_{\rm ion}$ measured in the outer
envelopes of molecular clouds (in a range of the gas column densities around $N\sim10^{21}$~cm$^{-2}$) varies significantly
from one object to another \citep[][]{Indriolo2012,Neufeld2017,Bacalla2019}. Given uncertainties in the leading transport
regime(s) governing the CR penetration into the clouds \citep[][]{Ivlev2018,Silsbee2019}, this introduces significant
uncertainty in the value of $\zeta_{\rm ion}$ near the center; available theories \citep[][]{Padovani2018} predict
$\zeta_{\rm ion}\sim10^{-17}-10^{-16}$~s$^{-1}$ for $N\sim10^{23}$~cm$^{-2}$.

The usual approach to calculate the gas and dust temperatures in dense cores \citep[see, e.g.,][]{Zucconi2001,Hocuk2017,
Chacon2019} relies on the assumption that the temperature of dust grains, $T_{\rm d}$, is determined from the balance of
their radiative heating (by absorbing far-IR interstellar radiation penetrating into the cores) and cooling (via the
continuum emission). Below we refer to this approach as the {\it standard model}, where the gas temperature $T_{\rm g}$ is
controlled by the CR heating and cooling on the dust surface, proportional to the difference $T_{\rm g}-T_{\rm d}$. The
available analytical models \citep[e.g.,][]{Goldsmith2001,Galli2002,Galli2015} usually make further simplification, assuming
the dust grains to be monodisperse, i.e., to all have the same size. Even though the term describing the gas-dust thermal
coupling is added to the dust energy balance in certain cases \citep[][]{Goldsmith2001,Woitke2009,Akimkin2013}, to the best
of our knowledge such calculations never account for an explicit dependence of the resulting dust temperature on the grain
size. At the same time, this dependence immediately follows from the very fact that the thermal coupling term is
proportional to the grain area, while the radiative absorption and emission terms scale with the grain volume in the
Rayleigh-Jeans regime.

The aim of the present paper is to develop a self-consistent analytical model for the equilibrium gas temperature and {\it
size-dependent} dust temperature in dense pre-stellar cores, depending on a given local gas density, local radiation field
and the CR ionization rate, and a power-law size distribution of dust grains. We derive compact expressions applicable for
typical conditions in dense cores, and compare our results with predictions of the commonly used standard model. Combined
with observations, our findings should have important implications -- in particular, for constraining the CR ionization rate
and the degree of dust evolution in dense cores, for discriminating between different regimes of CR transport in molecular
clouds, and for estimating the speed of physical and chemical processes occurring on the surface of grains.

\section{Heating and cooling in dense cores}
\label{heating-cooling}

In this section we summarize the main heating and cooling mechanisms of gas and dust, relevant to dense pre-stellar cores,
and formulate the respective balance equations.

\subsection{Balance equation for dust}
\label{BE_d}

The UV and visible radiation of the ISRF are practically completely attenuated in dense cores \citep[see, e.g.,][and
references therein]{Hocuk2017}. Therefore, dust is heated by absorbing photons from the far-IR part of the spectrum, still
able to penetrate into the cores.\footnote{As discussed by \citet[][]{Zucconi2001}, UV and optical radiation is converted
into far-IR radiation in a photodissociation layer surrounding the core. This could contribute to the far-IR heating, but
typically is not taken into account in radiative transfer models.} This is usually assumed to be the only heating mechanism,
which is balanced by the modified black-body radiative cooling of dust grains. In the present paper, we also include the
effect of the gas-dust thermal coupling. This leads to additional collisional heating due to gas particles impinging on the
dust surface.\footnote{The thermal coupling always leads to dust heating if it is the only mechanism of gas cooling (see
Section~\ref{BE_g}).} Then the energy balance for a dust grain can be written in the following form
\citep[][]{Goldsmith2001}:
\begin{equation}\label{balance_d}
\dot E_{\rm em}-\dot E_{\rm abs}=\dot E_{\rm gd},
\end{equation}
where $\dot E_{\rm em}$ and $\dot E_{\rm abs}$ are the rates of the radiative cooling (emission) and heating (absorption),
and $\dot E_{\rm gd}$ is the rate of collisional heating.

The rate of radiative heating of a grain of radius $a$ is \citep[][]{Draine2011Book}
\begin{equation*}
\dot E_{\rm abs} =\pi a^2 c\int u_\nu(N) Q_{\rm abs}(\nu,a)\:d\nu,
\end{equation*}
where $u_\nu(N)$ is the specific energy density of the ISRF at the gas column density $N$, $Q_{\rm abs}(\nu,a)$ is the dust
absorption efficiency for the frequency $\nu$, and $c$ is the speed of light. For the radiation dominated by far-IR photons
$Q_{\rm abs}(\nu,a)\propto \nu^2a$ (the spectral index of the dust opacity is assumed to be equal to 2). Hence, $E_{\rm
abs}=f(N)a^3$, where $f(N)$ is a function of $N$.

The rate of radiative cooling is \citep[][]{Draine2011Book}
\begin{equation*}
\dot E_{\rm em}=4\pi a^2\langle Q_{\rm abs}(\nu,a)\rangle_{T_{\rm d}}\sigma T_{\rm d}^4(a).
\end{equation*}
Here, $\langle Q_{\rm abs}\rangle_{T_{\rm d}}=q_{\rm abs}T_{\rm d}^2a$ is the Planck-averaged absorption efficiency of dust,
where $T_{\rm d}$ is the size-dependent dust temperature, $\sigma$ is the Stefan-Boltzmann constant, and $q_{\rm abs}$ is a
material-dependent numerical factor, equal to $\approx0.13$~K$^{-2}$~cm$^{-1}$ for silicate grains (we assume this value for
calculations below). We obtain
\begin{equation}\label{E_em}
\dot E_{\rm em}=4\pi q_{\rm abs}\sigma T_{\rm d}^6(a)a^3.
\end{equation}
Then the radiative heating rate can be conveniently presented in the following identical form:
\begin{equation}\label{E_abs}
\dot E_{\rm abs}\equiv4\pi q_{\rm abs}\sigma T_{\rm d0}^6(N)a^3,
\end{equation}
where $T_{\rm d0}$ is the local equilibrium dust temperature from the absorption-emission balance, i.e., when $\dot E_{\rm
gd}$ in Equation~(\ref{balance_d}) is neglected.

The gas-dust thermal coupling is characterized by the rate of collisional heating of a dust grain
\citep[][]{Draine2011Book,Burke1983},
\begin{equation}\label{E_gd}
\dot E_{\rm gd}=4\sqrt{2\pi}\:a^2\alpha_{\rm g}n_{\rm g}v_{\rm g}^*k_{\rm B}[T_{\rm g}-T_{\rm d}(a)],
\end{equation}
where $T_{\rm g}$ is the gas temperature, $v_{\rm g}^*$ is the thermal velocity scale of gas particles, $\alpha_{\rm g}$ is
their thermal accommodation coefficient on dust surface, and $k_{\rm B}$ is Boltzmann's constant. To include the effect of
the ISM elemental composition, we set the density of H$_2$ molecules as the relevant gas density scale $n_{\rm g}$. Then the
velocity scale,
\begin{equation*}
v_{\rm g}^*=\left(1+2\sqrt{2}\sum_i\frac{x_i}{\sqrt{\mu_i}}\right)v_{\rm g}\approx 1.14v_{\rm g},
\end{equation*}
is determined by the mass numbers $\mu_i$ of heavier elements and their abundances $x_i$ (with respect to the atomic
hydrogen), with $v_{\rm g}=\sqrt{k_{\rm B}T_{\rm g}/m_{\rm g}}$ being the thermal velocity scale of H$_2$ molecules. For the
accommodation coefficient, we adopt $\alpha_{\rm g}\approx0.5$ \citep[][]{Draine2011Book}.

\subsection{Balance equation for gas}
\label{BE_g}

The gas heating in pre-stellar cores is completely dominated by CRs, whereas the cooling occurs through two mechanisms
\citep[][]{Goldsmith2001,Galli2002}: Apart from the gas-dust thermal coupling, the molecular lines may contribute to the
cooling. The main line coolant in dense cores is the low-$J$ rotational transitions of CO molecules, whose catastrophic
freeze-out makes this mechanism unimportant for typical densities in the center of dense cores \citep[][]{Caselli1999,
Goldsmith2001}. Using the gas-grain chemical model by \citet[][]{Sipila2019} in conjunction with the hydrodynamical model by
\citet[][]{Sipila2018}, we have concluded that the line cooling can be safely neglected for $n_{\rm g}\gtrsim
10^5$~cm$^{-3}$, and then we can write the thermal balance as
\begin{equation}\label{balance_g}
\Gamma_{\rm CR}=\Lambda_{\rm gd}.
\end{equation}
Here, $\Gamma_{\rm CR}$ is the CR heating rate and $\Lambda_{\rm gd}$ is the cooling function due to gas-dust thermal
coupling.

The heating rate per unit volume can be presented in the following form \citep[][]{Glassgold2012}:
\begin{equation}\label{Gamma_CR}
\Gamma_{\rm CR}=\zeta_{\rm ion}(N)\varepsilon_{\rm heat}(n_{\rm g})n_{\rm g},
\end{equation}
where $\zeta_{\rm ion}$ is the total CR ionization rate per H$_2$ molecule and $\varepsilon_{\rm heat}$ is the heating
energy per H$_2$ ionization. The latter is a very slowly increasing function of $n_{\rm g}$, reaching values of $15-17$~eV
for dense cores; below we adopt the value of $\varepsilon_{\rm heat}=16$~eV.

The gas cooling function is determined by the rate of dust collisional heating,
\begin{equation}\label{Lambda_gd}
\Lambda_{\rm gd}=\int \dot E_{\rm gd}\:d n_{\rm d}.
\end{equation}
We generally assume an ``evolved MRN'' size distribution of grains,
\begin{equation}\label{MRN}
\frac{d n_{\rm d}}{da}\propto a^{-3.5+\gamma},
\end{equation}
in the range of $a_{\rm min}\leq a\leq a_{\rm max}$. The deviation of $a_{\rm min}$ and $a_{\rm max}$ from the ``canonical''
MRN values as well as nonzero $\gamma$ parameterize the degree of dust evolution due to possible coagulation in cores
\citep[][]{Weingartner2001b}. The value of $a_{\rm max}$ is not expected to exceed a few tenths of $\mu$m, while $a_{\rm
min}$ may increase significantly due to efficient depletion of smaller grains onto bigger ones. The slope variation is
usually positive and could be as large as $\gamma\approx1$ (or even larger). The scale factor in Equation~(\ref{MRN}) is
determined from the relation between the mass densities of dust and gas,
\begin{equation}\label{d_to_g}
f_{\rm d}^*m_{\rm g}n_{\rm g}=\frac43\pi\rho_{\rm d}\int_{a_{\rm min}}^{a_{\rm max}} a^3\frac{dn_{\rm d}}{da}\:da.
\end{equation}
Here, $\rho_{\rm d}$ is the mass density of grain material, equal to $\approx3.5$~g~cm$^{-3}$ for compact silicate, and
$f_{\rm d}^*$ is the dust-to-gas mass ratio where the normalization is by $m_{\rm g}n_{\rm g}$,
\begin{equation*}
f_{\rm d}^*=\left(1+\sum_ix_i\mu_i\right)f_{\rm d}\approx 1.4f_{\rm d},
\end{equation*}
expressed in terms of the standard value of $f_{\rm d}=0.01$. Below we adopt these values of $\rho_{\rm d}$ and $f_{\rm d}$.

We stress that the absorption term in Equation~(\ref{balance_d}) is completely determined by the local density of the ISRF,
i.e., the coupling to the local thermal radiation of dust is neglected. This approach is justified provided the optical
depth of the core is small for the thermal radiation. Using the above expression for $\langle Q_{\rm abs}\rangle_{T_{\rm
d}}$, we obtain $\approx \pi q_{\rm abs}T_{\rm d}^2 a^3$ for the corresponding absorption cross section. Then, utilizing
relation (\ref{d_to_g}) and the fact that the cross section scales with $a^3$, we readily infer the optical depth,
$\tau_{\rm core}\approx\frac34(q_{\rm abs}f_{\rm d}^*T_{\rm d}^2/\rho_{\rm d})\Sigma_{\rm core}$, where $\Sigma_{\rm
core}=m_{\rm g}N_{\rm core}$ is the H$_2$ surface density of the core. For cold dense cores with $T_{\rm d}\approx6$~K and
the column density of $N_{\rm core}\sim10^{23}$~cm$^{-2}$ \citep[][]{Crapsi2007,Keto2010} we estimate $\tau_{\rm
core}\sim3\times10^{-3}$. As the peak column density is expected to approximately scale with the square root of the peak
volume density \citep[for the Bonnor-Ebert sphere,][]{Bonnor1956}, our approach should be well applicable for $n_{\rm g}$ of
up to $\sim10^{10}$~cm$^{-3}$.

\subsection{Additional mechanisms of CR heating} \label{additional}

To facilitate the analysis and comparison with previous results, above we only discussed the heating included in the
standard model for the equilibrium gas and dust temperatures. Apart from this, there are additional mechanisms of dust and
gas heating by CRs, associated with the presence of dust grains. The dust heating is due to direct CR bombardment, leading
to the energy deposition into the grains \citep[e.g.,][]{Leger1985,Shen2004}, and due to absorption of UV radiation caused
by the CR-induced fluorescence of H$_2$ and He \citep[][]{Prasad1983,Cecchi-Pestellini1992}. Near-IR radiation from
vibrationally excited H$_2$ may add to dust heating \citep[][]{Dalgarno1999}, while the UV radiation also contributes to gas
heating via photoelectric emission from dust \citep[][]{Draine1978}. Furthermore, CRs are the only source of atomic hydrogen
in dense cores \citep[][]{Padovani2018b}. The energy released in the recombination on the surface of grains is distributed
between dust and gas.

The mechanisms of dust heating by CRs require a careful analysis, which is presented in Appendix~\ref{A1}. We show that for
typical conditions in dense pre-stellar cores (discussed in Section~\ref{example}) this heating is unimportant.

The effect of additional gas heating by CRs is straightforwardly included in Equations~(\ref{balance_g}) and
(\ref{Gamma_CR}), by adding the corresponding energy to $\varepsilon_{\rm heat}$. The upper bound for the energy of
photoelectric heating can be estimated as a product of the photoelectric yield and the energy of CR-induced UV heating of
dust. With the energy of $\approx 8$~eV \citep[][]{Dalgarno1999} and the yield of $\sim 0.1$ for Lyman-Werner photons and
MRN silicate grains \citep[][]{Weingartner2006}, we obtain the photoelectric heating energy of less than 1~eV per H$_2$
ionization. The gas heating due to hydrogen recombination, although uncertain, is expected to give a comparable contribution
\citep[][]{Glassgold2012}. As these energies are substantially smaller than the adopted value of $\varepsilon_{\rm heat}$,
we conclude that the additional gas heating can also be neglected.

\section{Equilibrium temperatures}
\label{eq_T}

Substituting Equations~(\ref{E_em})--(\ref{E_gd}) in Equation~(\ref{balance_d}), we get the thermal balance equation for
dust grains:
\begin{equation}\label{T_dust}
[T_{\rm d}^6(a)-T_{\rm d0}^6]a
=\sqrt{\frac2{\pi}}\:\frac{\alpha_{\rm g}n_{\rm g}v_{\rm g}^*}{q_{\rm abs}\sigma}\:k_{\rm B}[T_{\rm g}-T_{\rm d}(a)].
\end{equation}
The complementary balance equation for gas particles is obtained from Equation~(\ref{balance_g}): To derive $\Lambda_{\rm
gd}$, we insert Equation~(\ref{E_gd}) in Equation~(\ref{Lambda_gd}), and calculate the scale factor for the size
distribution (\ref{MRN}). Substituting the result in Equation~(\ref{balance_g}) and using Equation~(\ref{Gamma_CR}), we get
\begin{eqnarray}
\zeta_{\rm ion}=\sqrt{\frac{18}{\pi}}\:\frac{\alpha_{\rm g}f_{\rm d}^*m_{\rm g}n_{\rm g}v_{\rm g}^*}{\varepsilon_{\rm heat}\rho_{\rm d}}
\:\frac{0.5+\gamma}{a_{\rm max}^{0.5+\gamma}-a_{\rm min}^{0.5+\gamma}}\hspace{1.5cm}\label{T_gas}\\
\times\int_{a_{\rm min}}^{a_{\rm max}} k_{\rm B}[T_{\rm g}-T_{\rm d}(a)]\:a^{-1.5+\gamma}\:da. \nonumber
\end{eqnarray}
Equations~(\ref{T_dust}) and (\ref{T_gas}) yield the self-consistent solution for $T_{\rm d}(a)$ and $T_{\rm g}$ for given
local ionization rate and local conditions in the core.

Below we give the analytical approximation of Equations~(\ref{T_dust}) and (\ref{T_gas}), allowing us to derive the explicit
dependence of the dust temperature on the grain size and better understand the mechanism behind the observed behavior.

\subsection{Analytical approximation}
\label{approx}

Assume $\Delta T_{\rm d}(a)\equiv T_{\rm d}(a)-T_{\rm d0}$ is sufficiently small, so that the lhs of Equation~(\ref{T_dust})
can be expanded in a series over $\Delta T_{\rm d}$. Keeping up to quadratic terms, $\approx 6T_{\rm d0}^5\Delta T_{\rm
d}+15T_{\rm d0}^4\Delta T_{\rm d}^2$, ensures accurate results for $\Delta T_{\rm d}/T_{\rm d0}\lesssim2/5$ and leads to a
quadratic equation for $\Delta T_{\rm d}$. Keeping the same accuracy for $\Delta T_{\rm g}$, we obtain
\begin{equation}\label{Td_vs_Tg}
\Delta T_{\rm d}(a)\approx\left(1-\frac52\:\frac{a/A}{(1+a/A)^2}\frac{\Delta T_{\rm g}}{T_{\rm d0}}\right)
\frac{\Delta T_{\rm g}}{1+a/A}\,,
\end{equation}
where $\Delta T_{\rm g}\equiv T_{\rm g}-T_{\rm d0}$ and
\begin{equation}\label{A}
A=\frac1{\sqrt{18\pi}}\:\frac{\alpha_{\rm g}n_{\rm g}v_{\rm g}^*k_{\rm B}}{q_{\rm abs}\sigma T_{\rm d0}^5}\,,
\end{equation}
is the ``critical'' grain radius. For ``overcritical'' grains with $a\gg A$ we have $\Delta T_{\rm d}(a)\ll \Delta T_{\rm
g}$, i.e., their temperature is practically equal to $T_{\rm d0}$, as usually assumed; on the other hand, for grains with
$a\lesssim A$ the thermal coupling to gas dominates their energy balance and, hence, their temperature approaches $T_{\rm
g}$. As a consequence, the contribution of small grains to the gas cooling [determined by the integral in
Equation~(\ref{T_gas})] can be reduced drastically, which should lead to higher gas temperatures compared to the case where
$T_{\rm d}=T_{\rm d0}$ is assumed.

\begin{figure}[!ht]
\begin{center}
\resizebox{\hsize}{!}{\includegraphics{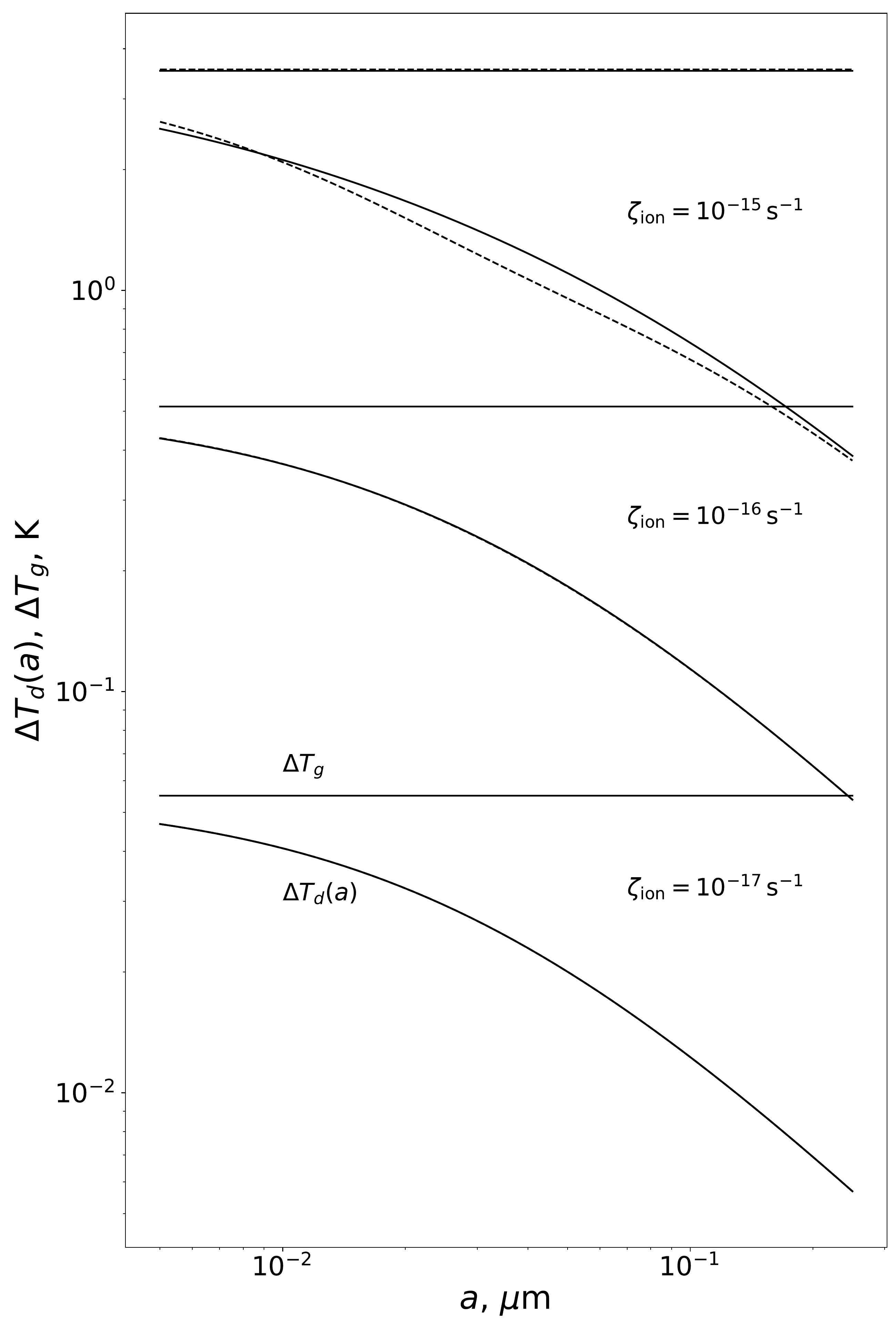}}
\caption{The excess of gas and dust temperature, $\Delta T_{\rm g}=T_{\rm g}-T_{\rm d0}$ and
$\Delta T_{\rm d}(a)=T_{\rm d}(a)-T_{\rm d0}$, above the temperature of the absorption-emission balance, $T_{\rm d0}$. The
three sets of solid lines illustrate the solution of Equations~(\ref{T_dust}) and (\ref{T_gas}) for the three indicated
values of the CR ionization rate $\zeta_{\rm ion}$, plotted versus the dust grain radius $a$ (the horizontal lines show
$\Delta T_{\rm g}$). The dashed lines (almost overlapped with the solid lines for $\Delta T_{\rm g}$) represent the
analytical approximation: Equations~(\ref{T_gas2}) and (\ref{Psi}) for $\Delta T_{\rm g}$ and Equation~(\ref{Td_vs_Tg}) for
$\Delta T_{\rm d}(a)$. The results are for $T_{\rm d0}=6$~K, assuming the canonical MRN size distribution ($\gamma=0$,
$a_{\rm min}=0.005~\mu$m, $a_{\rm max}=0.25~\mu$m) and the gas density $n_{\rm g}=10^6$~cm$^{-3}$.} \label{fig1}
\end{center}
\end{figure}

The critical radius is determined by the material-dependent absorption factor $q_{\rm abs}$ and by the local conditions in
the core (in particular, it depends on $T_{\rm g}$ via $v_{\rm g}^*$). Note that $A$ increases with $n_{\rm g}$ faster than
linearly, since $T_{\rm d0}$ is a decreasing function of $N$ \citep[][]{Evans2001,Hocuk2017,Chacon2019}.

We introduce dimensionless parameters
\begin{equation*}
\tilde R=a_{\rm max}/a_{\rm min}, \qquad \tilde A=A_0/a_{\rm min},
\end{equation*}
where $A_0\equiv A(T_{\rm g}=T_{\rm d0})$. Parameter $\tilde R$ quantifies the relative width of the size distribution,
while $\tilde A$ characterizes the relative importance of the critical radius: for $\tilde A\gtrsim1$ the overall effect of
finite $\Delta T_{\rm d}$ is expected to become significant. Substituting Equation~(\ref{Td_vs_Tg}) in
Equation~(\ref{T_gas}) and expanding $v_{\rm g}^*(T_{\rm g})$ in a series over $\Delta T_{\rm d}$, after some manipulation
we obtain:
\begin{equation}\label{T_gas2}
\zeta_{\rm ion}\approx\sqrt{\frac{18}{\pi}}\:\frac{\alpha_{\rm g}f_{\rm d}^*m_{\rm g}n_{\rm g}v_0^*k_{\rm B}T_{\rm d0}}
{\varepsilon_{\rm heat}\rho_{\rm d}a_{\rm min}}\:\Psi(\Delta T_{\rm g})\,,
\end{equation}
where $v_0^*\equiv v_{\rm g}^*(T_{\rm g}=T_{\rm d0})$. Function $\Psi(\Delta T_{\rm g})$ gives the explicit dependence on
the temperature difference as well as on the dimensionless parameters of the size distribution,
\begin{equation}\label{Psi}
\Psi=\frac{0.5+\gamma}{\tilde R^{0.5+\gamma}-1}\left(I_1+I_2\frac{\Delta T_{\rm g}}{T_{\rm d0}}\right)
\frac{\Delta T_{\rm g}}{T_{\rm d0}}\,,
\end{equation}
where auxiliary functions $I_{1,2}(\tilde A,\tilde R, \gamma)$ are determined by Equation~(\ref{I12}) in Appendix~\ref{A2}.
Thus, Equation~(\ref{T_gas2}) provides us with a convenient direct relation between the gas temperature and the ionization
rate.

In Figure~\ref{fig1} we compare the exact and analytical solutions. The solid lines represent the solution of
Equations~(\ref{T_dust}) and (\ref{T_gas}) for three different values of $\zeta_{\rm ion}$, showing $\Delta T_{\rm d}$ (as a
function of grain radius) and the corresponding $\Delta T_{\rm g}$. We set $T_{\rm d0}=6$~K for the illustration, which is
about the value expected from the absorption-emission balance in the center of very dense cores \citep[assuming a typical
ISRF, see, e.g.,][]{Crapsi2007}. The results are obtained for the canonical MRN distribution and a gas density typical to
such cores \citep[$n_{\rm g}=10^6$~cm$^{-3}$, see, e.g.,][]{Crapsi2005}, demonstrating that the temperature of smaller
grains approaches $T_{\rm g}$, while for bigger grains it approaches $T_{\rm d0}$. This trend is observed for all used
values of $\zeta_{\rm ion}$. According to Equation~(\ref{Td_vs_Tg}), the transition occurs at $a\approx A$, which is about
$0.03~\mu$m for the chosen conditions. As the critical radius $A$ scales (faster than) linearly with the gas density, an
increase in $n_{\rm g}$ by a factor of $\sim10$ will make $T_{\rm d}(a)$ almost equal to $T_{\rm g}$ for grains below
$a\sim0.1~\mu$m.

The analytical approximation is depicted in Figure~\ref{fig1} by the dashed lines. We see that Equations~(\ref{T_gas2}) and
(\ref{Psi}) provide excellent accuracy for the gas temperature for all values of $\zeta_{\rm ion}$ used in calculations, so
that the dashed and solid lines for $\Delta T_{\rm g}$ are undistinguishable. Some deviation of Equation~(\ref{Td_vs_Tg})
from the exact solution for $\Delta T_{\rm d}(a)$ (occurring at $a\sim A$) is only seen in the extreme case of $\zeta_{\rm
ion}=10^{-15}$~s$^{-1}$.

\section{Discussion}
\label{Discuss}

Let us quantify the difference between the standard model for gas temperature and our model for $T_{\rm g}$ and $T_{\rm
d}(a)$. To facilitate the analysis, we introduce a concept of {\it effective grain radius} and replace the self-consistent
dependence $\Delta T_{\rm d}(a)$, given by Equation~(\ref{T_dust}), with an effective (size-independent) value $\Delta
T_{\rm d,eff}$. The latter is derived by integrating Equation~(\ref{balance_d}) over the size distribution (\ref{MRN}) and
assuming that $T_{\rm d}$ does not depend on $a$. The results of this approach should tend to exact results both for small
$\tilde A$ (where the deviation of $T_{\rm d}$ from $T_{\rm d0}$ is insignificant) and for sufficiently large $\tilde A$
(where all grains in the MRN size range have the temperature close to $T_{\rm g}$).

\subsection{Effective grain radius}
\label{Discuss1}

Following the steps of Sec.~\ref{approx}, we obtain the relation between $\Delta T_{\rm d,eff}$ and $\Delta T_{\rm g}$ in
the form of Equation~(\ref{Td_vs_Tg}), where $a$ should be replaced with the effective grain radius $a_{\rm eff}=a_{\rm
min}\tilde a_{\rm eff}$. It is determined by
\begin{equation}\label{a_eff}
\tilde a_{\rm eff}(\tilde R,\gamma)=\frac{-0.5+\gamma}{\tilde R^{-0.5+\gamma}-1}\:\frac{\tilde R^{0.5+\gamma}-1}{0.5+\gamma}\,,
\end{equation}
a monotonically increasing function of both variables. For $\gamma\geq0$, it is bound between $\tilde R^{0.5}\leq\tilde
a_{\rm eff}<\tilde R$, i.e., $(a_{\rm min}a_{\rm max})^{0.5}\leq a_{\rm eff}<a_{\rm max}$. The lower bound corresponds to
$\gamma=0$; for $\gamma=0.5$ and $\gamma=1$ we have $\tilde a_{\rm eff}=(\tilde R-1)/\ln\tilde R$ and $\tilde a_{\rm
eff}=(\tilde R+\tilde R^{0.5}+1)/3$, respectively. The case of monodisperse grains is naturally recovered for $\tilde R\to
1$, where $a_{\rm eff}\to a_{\rm min}$. As discussed in Section~\ref{evolved}, the value of $a_{\rm eff}$ plays the critical
role in determining the value of $\Delta T_{\rm g}$.

The gas temperature is then derived from Equation~(\ref{T_gas2}) where $\Psi$ should be replaced with the corresponding
effective function,
\begin{eqnarray}
\Psi_{\rm eff}=\frac{1/\tilde a_{\rm eff}}{1+\tilde A/\tilde a_{\rm eff}} \hspace{5.5cm}\label{Psi_eff}\\
\times\left[1+\frac12\left(\frac1{1+\tilde A/\tilde a_{\rm eff}}+\frac5{(1+\tilde a_{\rm eff}/\tilde A)^2}\right)
\frac{\Delta T_{\rm g}}{T_{\rm d0}}\right]\frac{\Delta T_{\rm g}}{T_{\rm d0}}\,.\nonumber
\end{eqnarray}
Let us analyze Equation~(\ref{Psi_eff}). In the limit of small $A_0/a_{\rm eff}$ (where the deviation of the dust
temperature from $T_{\rm d0}$ is negligible), the factor in the brackets is $\approx\sqrt{T_{\rm g}/T_{\rm d0}}$ for the
accepted accuracy. This yields
\begin{equation}\label{Psi_small}
A_0\ll a_{\rm eff}:\quad \zeta_{\rm ion}\approx \sqrt{\frac{18}{\pi}}\:\frac{\alpha_{\rm g}f_{\rm d}^*m_{\rm g}n_{\rm g}}
{\varepsilon_{\rm heat}\rho_{\rm d}a_{\rm eff}}v_{\rm g}^*(T_{\rm g})k_{\rm B}\Delta T_{\rm g}\,.
\end{equation}
We see that Equation~(\ref{Psi_small}) represents the standard model for the gas temperature \citep[][]{Goldsmith2001,
Galli2002}, which is now {\it generalized} to the case of arbitrary power-law size distribution with the effective grain
radius $a_{\rm eff}$.

The absorption-emission balance assumed in the standard model is completely invalid for large $A_0/a_{\rm eff}$ (where the
effective dust temperature substantially exceeds $T_{\rm d0}$). The second line of Equation~(\ref{Psi_eff}) in this limit
represents the first two terms of expansion of $(T_{\rm g}/T_{\rm d0})^6-1$ over $\Delta T_{\rm g}$. With the same accuracy,
we can write\footnote{This expression can be readily obtained from Equation~(\ref{balance_d}) by integrating the latter over
the size distribution and equating the resulting rhs to $\Gamma_{\rm CR}$. Then, assuming $T_{\rm d}\approx T_{\rm g}$ and
utilizing relation (\ref{d_to_g}), we arrive to Equation~(\ref{T_gas3}).}
\begin{equation}\label{T_gas3}
A_0\gg a_{\rm eff}:\quad \zeta_{\rm ion}\approx 3\:\frac{f_{\rm d}^*m_{\rm g}q_{\rm abs}\sigma}
{\varepsilon_{\rm heat}\rho_{\rm d}}\left(T_{\rm g}^6-T_{\rm d0}^6\right).
\end{equation}
For sufficiently small $\Delta T_{\rm g}$, the rhs of Equation~(\ref{T_gas3}) is a factor of $1+A_0/a_{\rm eff}$ smaller
than that of Equation~(\ref{Psi_small}), as it follows from Equation~(\ref{Psi_eff}). This implies that $\Delta T_{\rm g}$
for $A_0\gg a_{\rm eff}$ is by this factor larger than the prediction of the standard theory. Furthermore, the nonlinearity
in Equation~(\ref{T_gas3}) is substantially stronger than that in Equation~(\ref{Psi_small}), which is another consequence
of the deviation of dust temperature from $T_{\rm d0}$ (neglected in the standard model). A remarkable fact is that
Equation~(\ref{T_gas3}) is explicitly independent of the gas density and characteristics of the size distribution [an
implicit dependence on $n_{\rm g}$ is via $\zeta_{\rm ion}(N)$]. This is a natural consequence of strong gas-dust thermal
coupling in the limit $A_0\gg a_{\rm eff}$, so $\Delta T_{\rm g}$ is a nearly universal function of the column density.

Since the critical grain radius, Equation~(\ref{A}), is an increasing function of $n_{\rm g}$ (and $N$), a transition to
large $A_0/a_{\rm eff}$ occurs at a certain critical density. For the MRN size distribution, Equation~(\ref{T_gas3}) is
applicable for $n_{\rm g}$ satisfying the condition $A_0(n_{\rm g})\gg(a_{\rm min}a_{\rm max})^{0.5}$, which requires gas
densities substantially larger than $10^6$~cm$^{-3}$.

\begin{figure}[!ht]
\begin{center}
\resizebox{\hsize}{!}{\includegraphics{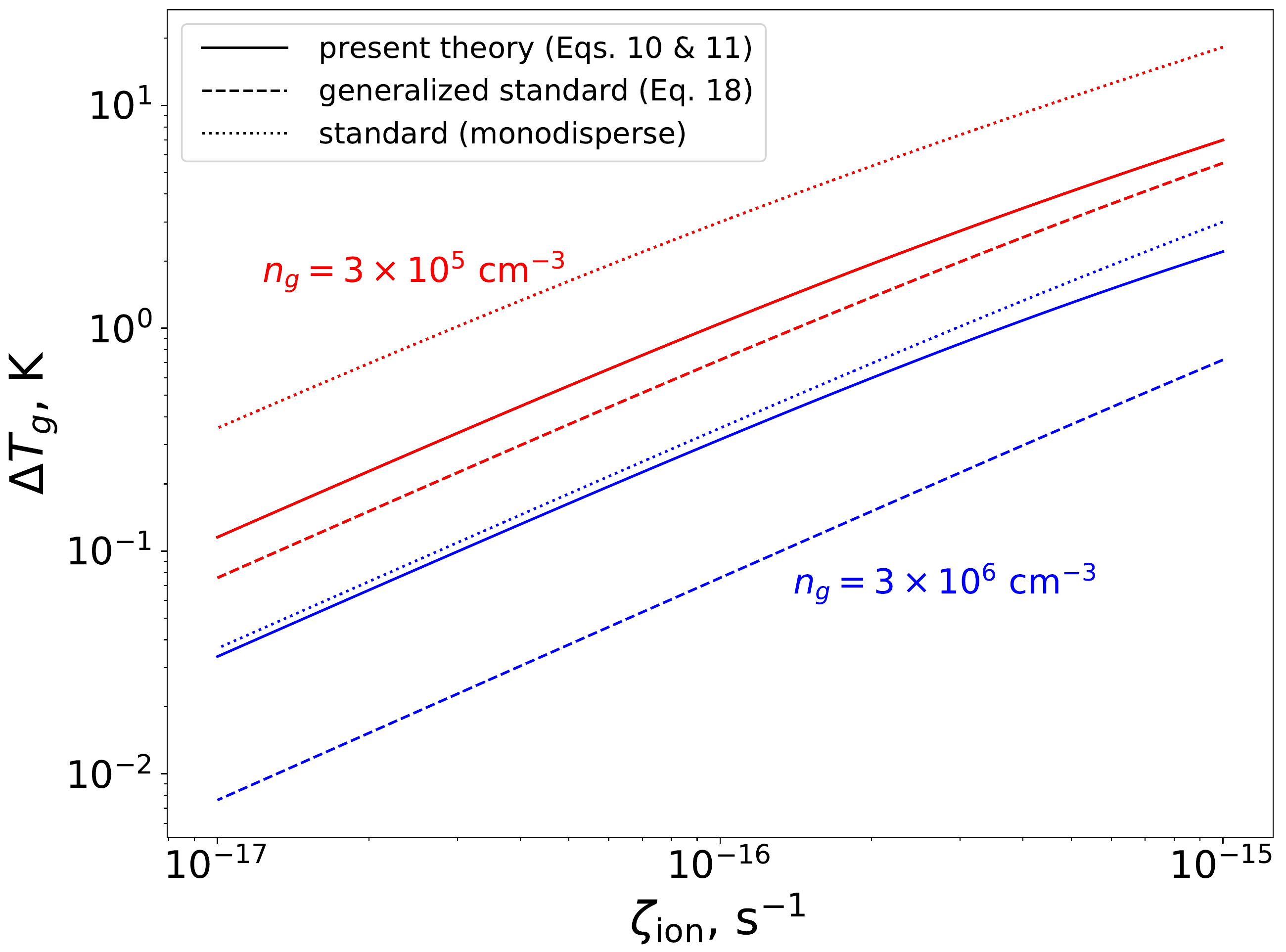}}
\caption{Gas temperature difference $\Delta T_{\rm g}$ as a function of the CR ionization rate $\zeta_{\rm ion}$. The solid
and dashed lines represent, respectively, the present theory (exact solution) and the prediction of the {\it generalized}
standard theory, both derived for the MRN size distribution. The dotted line shows $\Delta T_{\rm g}$ calculated from the
standard theory assuming {\it monodisperse} grains of $a=0.17~\mu$m. The upper and lower sets of curves are for $n_{\rm
g}=3\times10^5$~cm$^{-3}$ and $3\times10^6$~cm$^{-3}$, respectively, the other conditions as in Figure~\ref{fig1}.}
\label{fig2}
\end{center}
\end{figure}

Figure~\ref{fig2} summarizes results for the gas temperature, presenting $\Delta T_{\rm g}(\zeta_{\rm ion})$ for $n_{\rm
g}=3\times10^5$~cm$^{-3}$ and $3\times10^6$~cm$^{-3}$. With the solid lines we plot the exact solution of
Equations~(\ref{T_dust}) and (\ref{T_gas}) [the analytical solution, Equations~(\ref{T_gas2}) and (\ref{Psi}), is nearly
indistinguishable], while the dashed lines depict the generalized standard model described by Equation~(\ref{Psi_small}). We
see that the latter systematically underestimates $\Delta T_{\rm g}$. The disparity increases with gas density, as $\approx
1+A_0(n_{\rm g})/a_{\rm eff}$: the exact solution at higher $n_{\rm g}$ approaches the value given by
Equation~(\ref{T_gas3}), whereas $\Delta T_{\rm g}$ predicted by the standard model keeps decreasing as $\propto n_{\rm
g}^{-1}$.

The dotted lines in Figure~\ref{fig2} highlight the crucial role of the size distribution, showing $\Delta T_{\rm
g}(\zeta_{\rm ion})$ derived from the standard model where grains are assumed to be monodisperse. Following the original
work of \citet[][]{Goldsmith2001}, we set the grain radius to the fiducial value of $a=0.17~\mu$m (while the other
parameters entering Equation~(\ref{Psi_small}) are kept the same as above). Since this value is a factor of a few larger
than $a_{\rm eff}=0.035~\mu$m for the MRN dust, for $n_{\rm g}\lesssim\times10^6$~cm$^{-3}$ the resulting $\Delta T_{\rm g}$
is systematically overestimated and the dotted line lies well above the solid line (while near $n_{\rm
g}\sim3\times10^6$~cm$^{-3}$ the discrepancy between the exact and the standard models are almost compensated). In the
following section we elaborate on the major effect of the size distribution.

\subsection{Impact of dust evolution}
\label{evolved}

Possible dust evolution due to ongoing coagulation of grains \citep[see, e.g.,][]{Flower2005,Chacon2017} should reduce the
total surface area of dust and, hence, the thermal coupling with gas, thus increasing $\Delta T_{\rm g}$. A reduction of the
area is described by the effective grain radius, Equation~(\ref{a_eff}), which is proportional to the dust mass-to-area
ratio. As we demonstrated in the beginning of Section~\ref{Discuss1}, $a_{\rm eff}$ is a rapidly increasing function of
$a_{\rm min}$, $a_{\rm max}$, and $\gamma$.

\begin{figure}[!ht]
\begin{center}
\resizebox{\hsize}{!}{\includegraphics{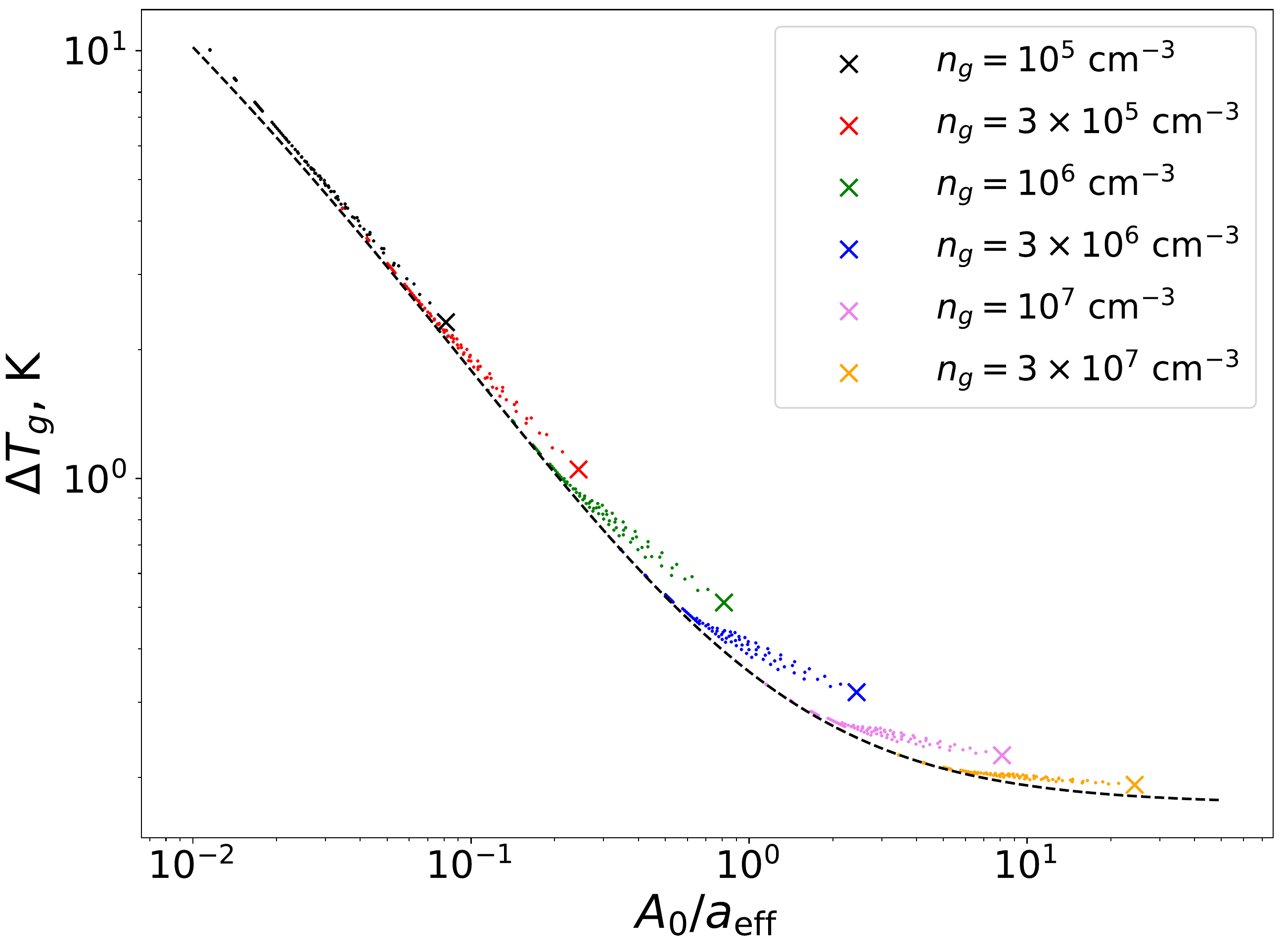}}
\caption{Gas temperature difference $\Delta T_{\rm g}$ versus the ratio $A_0(n_{\rm g})/a_{\rm eff}$. The exact results,
Equations~(\ref{T_dust}) and (\ref{T_gas}), are plotted for $\zeta_{\rm ion}=10^{-16}$~s$^{-1}$ and for six characteristic
values of $n_{\rm g}$ (color-coded, see the inset). The crosses indicate the non-evolved MRN dust; the dust evolution is
quantified by the value of $a_{\rm eff}$, calculated for varying $\gamma$ and $a_{\rm min}$ while $a_{\rm max}$ is fixed
(see text for details). The dashed line represents the approach of effective grain radius,
Equations~(\ref{T_gas2}) and (\ref{Psi_eff}).} \label{fig3}
\end{center}
\end{figure}

To describe the impact of dust evolution on the gas temperature and investigate how accurately the effective radius
quantifies this, in Figure~\ref{fig3} we plot $\Delta T_{\rm g}$ as a function of the ratio $A_0/a_{\rm eff}$. We consider a
range of gas densities of $10^5$~cm$^{-3}\leq n_{\rm g}\leq3\times10^7$~cm$^{-3}$, representative of very dense pre-stellar
cores \citep[][]{Keto2010,Caselli2019}. This gives the range of the critical radii $A_0(n_{\rm g})$, as determined by
Equation~(\ref{A}) for $T_{\rm g}=T_{\rm d0}$. Assuming the evolution primarily leads to depletion of smaller grains, we
vary $a_{\rm min}$ in the whole size range of the canonical MRN dust, $0.005~\mu$m~$\leq a_{\rm min}\leq0.25~\mu$m, while
$a_{\rm max}$ is fixed to the maximum value and $\gamma$ varies between 0 and 1 \citep[][]{Weingartner2001b}. This yields
$a_{\rm eff}$ varying between $0.035~\mu$m (non-evolved MRN dust) and $0.25~\mu$m (highly-evolved dust, concentrated at the
upper bound of the MRN distribution). The exact values of $\Delta T_{\rm g}$ (color-coded dots) are then computed from
Equations~(\ref{T_dust}) and (\ref{T_gas}). The approach of effective grain radius, Equations~(\ref{T_gas2}) and
(\ref{Psi_eff}), is represented by the dashed line.

Figure~\ref{fig3} shows that for sufficiently low gas densities, $n_{\rm g}\lesssim3\times10^5$~cm$^{-3}$, the resulting
values of $A_0/a_{\rm eff}$ are small. Then the exact $\Delta T_{\rm g}$ does not practically depend on the degree of the
dust evolution and tends to the prediction of the generalized standard model -- the limit described by
Equation~(\ref{Psi_small}). The parameters of the size distribution have a fairly weak impact on $\Delta T_{\rm g}$ also at
high densities, $n_{\rm g}\gtrsim10^7$~cm$^{-3}$, where $A_0/a_{\rm eff}$ is large, and the plot tends to the universal
asymptote of Equation~(\ref{T_gas3}). We note that for $\Delta T_{\rm g}\lesssim 3$~K, where the nonlinearity is negligible,
the plot simply scales with $\zeta_{\rm ion}$.

Figure~\ref{fig3} also highlights the fact that the maximum discrepancy between the approach of effective grain radius and
the exact results occurs near $A_0/a_{\rm eff}\sim1$. The discrepancy does not exceed 30\% for the MRN dust and is naturally
reduced for evolved dust, since Equation~(\ref{Psi_eff}) becomes exact for monodisperse grains. As expected, in the limits
of small or large $A_0/a_{\rm eff}$ the exact results converge to the dashed line.\footnote{A marginal deviation of the
dashed line from the exact results seen in Figure~\ref{fig3} for $\Delta T_{\rm g}\gtrsim3$~K is due to a series expansion
of $v_{\rm g}^*(T_{\rm g})$, used to derive Equation~(\ref{Psi_eff}). In this case, the dependence is perfectly described by
Equation~(\ref{Psi_small}).}

We recall that evolved dust is comprised of aggregates of smaller solid grains, and thus the effective material density of
such aggregates is expected to decrease with size. The effective density can be described by adopting a common approach of
fractal dust \citep[see, e.g.,][and references therein]{Okuzumi2009b}, where the mass of a particle of radius $a$ is assumed
to scale as $\propto a^D$, with $2\lesssim D \lesssim3$ being the fractal dimension. While the discussion of existing
fractal models and the choice of appropriate fractal dimension is beyond the scope of the present paper, we note that our
results can, in principle, be generalized for fractal dust, by substituting the corresponding mass scaling into the rhs of
Equation~(\ref{d_to_g}).

\subsection{Dust emission}
\label{emission}

The continuum dust emission is another observable to characterize the processes occurring in dense cores
\citep[][]{Chacon2019}. The emission is given by Equation~(\ref{E_em}) averaged over the grain size distribution. The
effective temperature of the emission can thus be defined as
\begin{equation}\label{T_deff0}
T_{\rm d,eff}^6=\frac{0.5+\gamma}{\tilde R^{0.5+\gamma}-1}\int_1^{\tilde R}T_{\rm d}^6(x)\:x^{-0.5+\gamma}\:dx,
\end{equation}
where $x=a/a_{\rm min}$. We obtain $T_{\rm d,eff}$ from the energy balance for dust grains, by integrating
Equation~(\ref{balance_d}) over the size distribution, equating the resulting rhs to $\Gamma_{\rm CR}$, and utilizing
relation (\ref{d_to_g}):
\begin{equation}\label{T_deff}
T_{\rm d,eff}^6=T_{\rm d0}^6+\frac{\zeta_{\rm ion}\varepsilon_{\rm heat}\rho_{\rm d}}{3f_{\rm d}^*m_{\rm g}q_{\rm abs}\sigma}\,.
\end{equation}
One can see that Equation~(\ref{T_deff}) is an inversion of Equation~(\ref{T_gas3}) with $T_{\rm d,eff}$ substituted for
$T_{\rm g}$, and therefore the two temperatures coincide in the limit of large $A_0/a_{\rm eff}$. We stress, however, that
Equation~(\ref{T_gas3}) is only applicable in the limit of strong gas-dust coupling, while Equation~(\ref{T_deff}) is
appropriate for all conditions where the governing equations of our model are valid. A notable property of $T_{\rm d,eff}$
is that it depends neither on gas density nor on the grain size distribution.

To explore how accurately the effective dust temperature represents the emission for different values of $\zeta_{\rm ion}$,
we have computed the total emissivity of grains for varying $n_{\rm g}$ and parameters of the size distribution (similar to
that in Figure~\ref{fig3}) and compared the results with the modified black-body radiation of grains at $T_{\rm d}=T_{\rm
d,eff}(\zeta_{\rm ion})$. We have found that the deviation of the effective spectral energy distribution from the exact
dependence in the Rayleigh-Jeans regime is practically negligible for $\zeta_{\rm ion}\lesssim10^{-15}$~s$^{-1}$; a weak
dependence on $n_{\rm g}$ is only seen far in the Planck regime (where the emissivity is already decreased by orders of
magnitude). As the higher values of $\zeta_{\rm ion}$ are unlikely in the local cores \citep[][]{Neufeld2017}, we conclude
that the effective temperature determined by Equation~(\ref{T_deff}) provides an excellent universal parametrization of the
dust emission.

\subsection{Example: The pre-stellar core L1544}
\label{example}

The physical structure of the pre-stellar core L1544 has been studied in detail in many publications \citep[][]{Tafalla2002,
Crapsi2007, Keto2015,Chacon2019,Caselli2019}. We have therefore chosen L1544 to illustrate the results of the present
theory, and to compare these with the predictions of the commonly used standard theory.

Depending on the particular model of dust opacity used by different authors for calculating the core structure, the dust
temperature ($T_{\rm d0}$) in the core center may vary between 6 and 7~K, while the peak gas density is predicted to be
between a few times $10^6$~cm$^{-3}$ and about $10^7$~cm$^{-3}$. We adopt the radial distributions for $n_{\rm g}$ and
$T_{\rm d0}$ from \citet[][]{Keto2015}, with the peak density of $8.3\times10^6$~cm$^{-3}$ and the central dust temperature
of 6.3~K (within a radius of 125~au), and calculate the dependencies $T_{\rm d0}(n_{\rm g})$ and $N(n_{\rm g})$. Then we
obtain $\zeta_{\rm ion}(n_{\rm g})$ from $\zeta_{\rm ion}(N)$ derived by \citet[][]{Padovani2018} (their model
$\mathscr{H}$), which yields the ionization rate around $10^{-16}$~s$^{-1}$ for $10^5$~cm$^{-3}\lesssim n_{\rm
g}\lesssim10^7$~cm$^{-3}$.

\begin{figure}[!ht]
\begin{center}
\resizebox{\hsize}{!}{\includegraphics{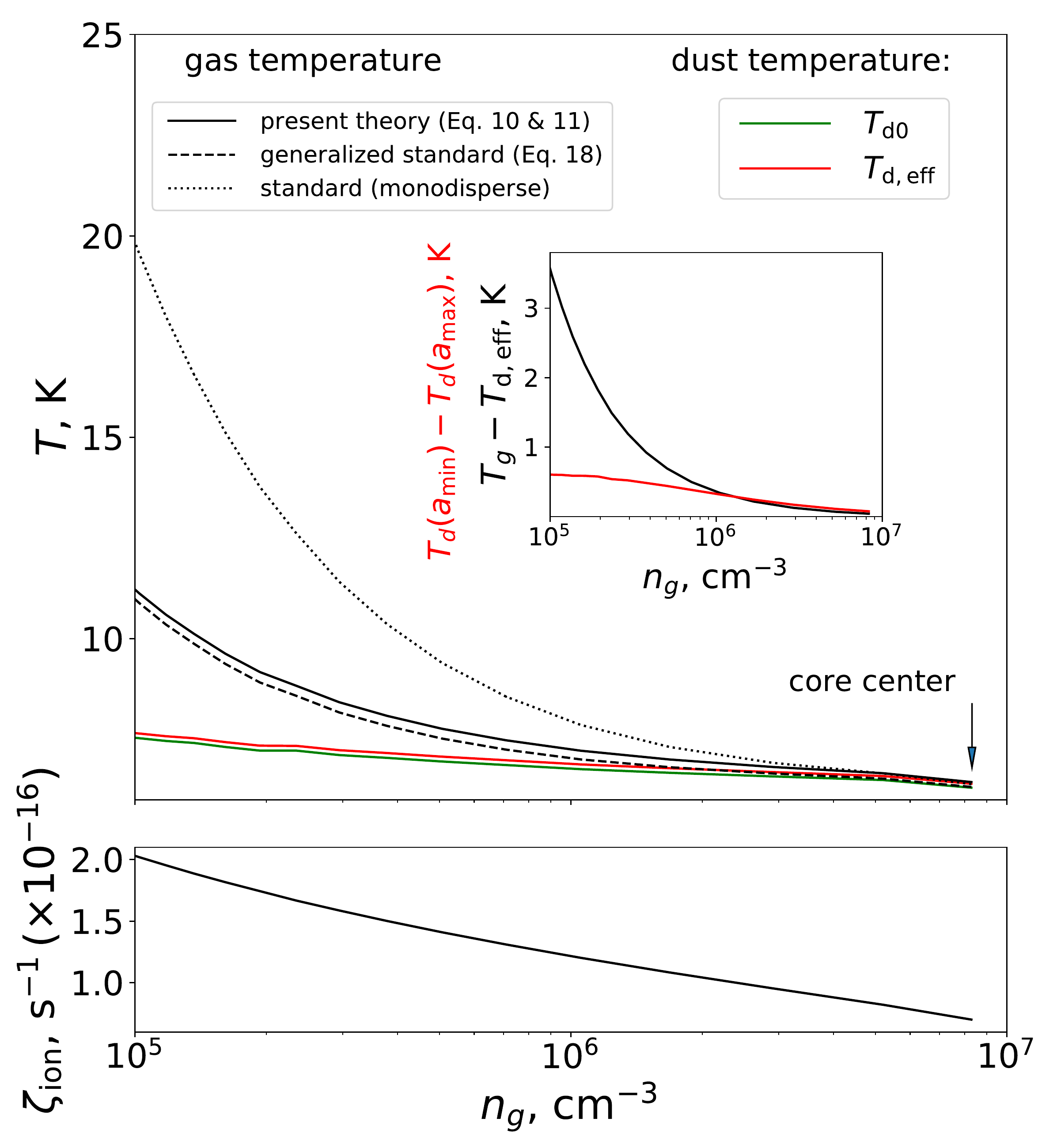}}
\caption{Upper panel: Dependence of the gas and dust temperature on the gas volume density $n_{\rm g}$ in the pre-stellar
core L1544. The non-evolved MRN distribution of grain sizes is assumed. For the gas temperature, the legend is the same as
in Figure~\ref{fig2}; for the dust temperature, the green and red lines represent $T_{\rm d0}$ and $T_{\rm d,eff}$,
respectively. In the inset, we plot the difference between $T_{\rm g}$ and $T_{\rm d,eff}$ (black line) as well as between
the temperatures of the smallest and largest grains (red line), both derived from the present theory. Lower panel: The CR
ionization rate $\zeta_{\rm ion}$ versus $n_{\rm g}$, calculated for L1544 from \citet[][]{Padovani2018}.} \label{fig4}
\end{center}
\end{figure}

Figure~\ref{fig4} summarizes results for the gas and dust temperatures (upper panel) as well as for the CR ionization rate
(lower panel), plotted versus the gas density in L1544. The black lines in the upper panel represent different models for
$T_{\rm g}$, calculated for the non-evolved MRN dust. As in Figure~\ref{fig2}, we compare the results of our exact theory,
Equations~(\ref{T_dust}) and (\ref{T_gas}), with the generalized standard model, Equation~(\ref{Psi_small}), and the
standard model for monodisperse grains. We see that the generalized standard model (dashed line) provides a fair
description, showing a nearly constant deviation of about 0.25~K from the exact results (solid line). The reason can be
directly understood from Figure~\ref{fig2}, where the temperature difference between the solid and dashed lines is, indeed,
about that value for $\zeta_{\rm ion}\sim10^{-16}$~s$^{-1}$. On the other hand, the standard model by
\citet[][]{Goldsmith2001,Galli2002} (dotted line), assuming all grains to have a certain fiducial size (which is
significantly larger than $a_{\rm eff}$ for the MRN dust, see Section~\ref{Discuss1}), predicts a gas temperature that is
substantially higher than the results of the present theory up to $n_{\rm g}\sim3\times 10^6$~cm$^{-3}$. At higher
densities, the dotted line goes below the solid line, following the trend seen in Figure~\ref{fig2}.

Unlike the gas temperature, the effective grain temperature $T_{\rm d,eff}$ characterizing the continuum emission (see
Section~\ref{emission}) does not depend on the size distribution. Its deviation from $T_{\rm d0}$ remains very small (about
$0.1$~K).

We conclude that for pre-stellar cores in the local ISM, such as L1544, the temperature difference $T_{\rm g}-T_{\rm d,eff}$
is practically equal to the difference $\Delta T_{\rm g}$, and the latter is reasonably described by the generalized
standard theory, Equation~(\ref{Psi_small}). The inset in Figure~\ref{fig4} suggests that its magnitude (black solid line)
is expected to be between $\approx3.5$~K and $\approx0.4$~K for $10^5$~cm$^{-3}\lesssim n_{\rm g}\lesssim10^6$~cm$^{-3}$,
i.e., should be measurable in this density range ($T_{\rm g}-T_{\rm d,eff}$ should start decreasing at lower densities,
where the line emission becomes the efficient mechanism of gas cooling). Hence, the temperature difference deduced from
observational data yields the product $\zeta_{\rm ion}a_{\rm eff}$, which should allow us in the future to constrain models
of both the CR ionization and dust evolution in dense cores.

We point out that the present theory, including the generalized standard model, yields gas temperatures in the central
region of L1544 which are very close to the measurements \citep[see, e.g., Figures~4(a) and 5 in][for $n_{\rm
g}\gtrsim10^5$~cm$^{-3}$]{Crapsi2007}. A notable fact is that our results are obtained assuming the non-evolved MRN
distribution of grains and using the ionization rate from \citet[][]{Padovani2018}, who suggest $\zeta_{\rm ion}\sim
10^{-16}$~s$^{-1}$ for this region, whereas the standard theory (monodisperse grains with $a_{\rm eff}\sim0.1~\mu$m)
requires $\zeta_{\rm ion}\sim 10^{-17}$~s$^{-1}$ to provide agreement with the measurements
\citep[e.g.,][]{Crapsi2007,Keto2010,Galli2015}. This reflects the crucial role of the grain size distribution (i.e., of a
proper choice for $a_{\rm eff}$).

The inset in Figure~\ref{fig4} also demonstrates the difference between the temperatures of the smallest and largest grains,
$T_{\rm d}(a_{\rm min})-T_{\rm d}(a_{\rm max})$ (red solid line). As follows from Equation~(\ref{Td_vs_Tg}), the difference
tends to $\approx\Delta T_{\rm g}$ when the critical radius $A_0$ exceeds $a_{\rm min}$ (the surface CR heating increases it
by the value of $T_{\rm g,s}\,$, see Equations~(\ref{shift_s}) and (\ref{T_gs}) in Appendix~\ref{A1}). For L1544, this
occurs at $n_{\rm g}\gtrsim4\times10^5$~cm$^{-3}$. At lower densities $T_{\rm d}(a_{\rm min})-T_{\rm d}(a_{\rm max})$
becomes significantly smaller than $\Delta T_{\rm g}$; its magnitude remains almost constant, reaching $\approx0.6$~K at
$n_{\rm g}=10^5$~cm$^{-3}$. This relatively small difference may, nevertheless, have important consequences for the physical
and chemical processes occurring on the surface of grains in cold dense cores, due to their extreme temperature dependence.

\section{Summary and outlook}
\label{Conclude}

In this paper we showed that the gas temperature $T_{\rm g}$ in dense pre-stellar cores strongly depends on parameters of
the grain size distribution. Furthermore, at high gas densities the value of $T_{\rm g}$ can substantially exceed the
predictions of the standard theory, in which the dust temperature $T_{\rm d}$ is assumed to be size-independent and equal to
$T_{\rm d0}(N)$ -- the value determined from the balance of radiative heating and cooling of a grain at the column density
$N$. Equations~(\ref{T_gas2}) and (\ref{Psi}) yield an accurate analytical relation between $T_{\rm g}$ and the CR
ionization rate $\zeta_{\rm ion}(N)$ for a given size distribution, valid for gas densities $n_{\rm g}\gtrsim10^5$~cm$^{-3}$
(where the gas cooling due to molecular line emission is negligible) and $\lesssim10^{10}$~cm$^{-3}$ (where the coupling to
the local thermal radiation of dust can be safely neglected).

We also derived an expression for the effective dust temperature $T_{\rm d,eff}$, Equation~(\ref{T_deff}), which provides a
parametrization of the spectral energy distribution of the continuum emission (valid for the same range of $n_{\rm g}$). We
found that the dust emissivity does not depend on the gas density or the grain size distribution, and is solely determined
by the values of $\zeta_{\rm ion}$ and $T_{\rm d0}$.

To facilitate the use of our model, in Appendix~\ref{A3} we present a convenient parametrization for the gas and dust
temperatures, applicable for typical conditions in dense cores:
\begin{enumerate}
\item {\it Gas temperature.} Equation~(\ref{rel1}) gives the relation between $T_{\rm g}$ and $\zeta_{\rm ion}$, and
    also depends on the local parameters $n_{\rm g}$ and $T_{\rm d0}$ as well as on $\tilde R$, the relative width of
    the size distribution. In particular, for $\zeta_{\rm ion}= 10^{-16}$~s$^{-1}$, $T_{\rm d0}=6$~K, and values of
    $n_{\rm g}$ chosen for the plot in Figure~\ref{fig3}, this relation yields the median curves through the
    corresponding color-coded points (varying $\gamma$ only leads to a slight scatter of the points off the curves, and
    therefore has a minor effect on the results). The curves connect the crosses (MRN dust) with the dashed line
    (monodisperse grains) in that figure, thus parameterizing the dependence of $\Delta T_{\rm g}$ on the size
    distribution.
\item {\it Effective dust temperature.} Equation~(\ref{rel2}), relating $T_{\rm d,eff}$ and $\zeta_{\rm ion}$, depends
    only on $T_{\rm d0}$. For $\zeta_{\rm ion}= 10^{-16}$~s$^{-1}$ and $T_{\rm d0}=6$~K, this yields $T_{\rm
    d,eff}-T_{\rm d0}\approx0.2$~K; as expected, this value coincides with the asymptotic value of $\Delta T_{\rm g}$ in
    the limit of large $A_0/a_{\rm eff}$ in Figure~\ref{fig3}.
\end{enumerate}

Our findings imply that measuring both $T_{\rm g}$ and $T_{\rm d,eff}$ in a range of gas densities between $10^5$~cm$^{-3}
\lesssim n_{\rm g}\lesssim10^6$~cm$^{-3}$ (where the predicted magnitude of $\Delta T_{\rm g}$ is expected to be above the
measurement uncertainty, see Figure~\ref{fig4}), one could develop a method for estimating and constraining the ionization
rate and the degree of dust evolution. The present theory predicts the values of $T_{\rm g}$ in the pre-stellar core L1544
which are very close to the measured values \citep[][]{Crapsi2007}, assuming the non-evolved MRN distribution of grain sizes
and the ionization rate as high as $\zeta_{\rm ion}\sim 10^{-16}$~s$^{-1}$. In the future we plan to carry out a detailed
analysis and check if our results could reproduce the ALMA and JVLA observations of \citet[][]{Caselli2019} and
\citet[][]{Crapsi2007} toward L1544. In general, by combining our model with models for initial stages of dust coagulation
in dense pre-stellar cores \citep[][]{Flower2005,Chacon2017}, we will gain deeper insights into fundamental physical
processes occurring in these objects and better understand the mechanisms controlling CR penetration into the clouds. For
example, comparing the CR ionization in diffuse and dense regions of molecular clouds should allow us to discriminate
between different transport regimes of CRs, resulting in largely different attenuation of $\zeta_{\rm ion}$ with the column
density \citep[][]{Silsbee2019}.

Finally, the conclusion that the temperature of smaller grains approaches $T_{\rm g}$ at higher gas densities, while bigger
grains are at $T_{\rm d}\approx T_{\rm d0}$ may have a profound impact on the speed of physical and chemical processes
occurring on the dust surface. The thermally activated desorption of atoms and molecules from the surface as well as the
surface diffusion and, hence, the diffusion-limited chemical reactions obey the Arrhenius temperature dependence, with the
typical activation energy of the order of hundreds of Kelvin \citep[e.g.,][]{Vasyunin2017}. Therefore, the temperature
increase by only a few tenths of Kelvin for smaller grains (dominating the dust surface) could lead to a significant
acceleration of these processes. Furthermore, given a highly uncertain ``average'' rate of the CR ionization in the ISM
\citep[][]{Indriolo2012}, and the fact that the ionization can be very strongly enhanced close to protostars
\citep[][]{Ceccarelli2014,Podio2014} and supernova remnants \citep[][]{Vaupre2014} -- serving as sources of the local CRs,
the surface chemistry could be affected dramatically. The additional dust heating by CRs, discussed in Appendix~\ref{A1},
could play especially important role in these environments. This problem will be studied in a separate paper.

We would like to thank Daniele Galli for useful discussions and suggestions, and an anonymous referee for constructive and
stimulating suggestions.

\appendix

\section{Appendix A\\ Additional dust heating by CRs}
\label{A1}

The CR heating of dust contains both surface (UV absorption, recombination) and volume (IR absorption, CR bombardment)
terms. Hence, a sum $\tau a^2+\upsilon a^3$ should be added to the rhs of Equation~(\ref{balance_d}), generalizing the
energy balance for a grain. The numerical factors are conveniently determined from
\begin{eqnarray*}
  \tau\int a^2\:dn_{\rm d}&=& \zeta_{\rm ion}\varepsilon_{\rm s}n_{\rm g}\,,\\
  \upsilon\int a^3\:dn_{\rm d}&=& \zeta_{\rm ion}\varepsilon_{\rm v}n_{\rm g}\,,
\end{eqnarray*}
where $\varepsilon_{\rm s,v}$ are the energies per H$_2$ ionization for the surface and volume heating, respectively (see
below). We immediately infer that the governing relation between the gas and dust temperatures, Equation~(\ref{T_dust}),
remains unchanged after the following replacement:
\begin{eqnarray}
  T_{\rm g}&\to&T_{\rm g}+T_{\rm g,s}\,, \label{shift_s}\\
  T_{\rm d0}^6&\to&T_{\rm d0}^6+T_{\rm dv}^6\,.\label{shift_v}
\end{eqnarray}
The respective ``shifts'' due to the surface and volume heating is then obtained from Equation~(\ref{T_dust}),
\begin{eqnarray}
  T_{\rm g,s}&=&\sqrt{\frac{\pi}{18}}\:\frac{\zeta_{\rm ion}\varepsilon_{\rm s}\rho_{\rm d}a_{\rm eff}}
  {\alpha_{\rm g}f_{\rm d}^*m_{\rm g}n_{\rm g}v_{\rm g}^*k_{\rm B}}\,, \label{T_gs}\\
  T_{\rm d,v}^6&=&\frac{\zeta_{\rm ion}\varepsilon_{\rm v}\rho_{\rm d}}{3f_{\rm d}^*m_{\rm g}q_{\rm abs}\sigma}\,, \label{T_dv}
\end{eqnarray}
where $a_{\rm eff}$ is the effective grain radius, Equation~(\ref{a_eff}). Replacing $T_{\rm g}$ with its shifted value in
Equation~(\ref{T_gas}), we conclude that this governing equation remains unchanged, too, if the ionization rate is replaced
with
\begin{equation}\label{zeta_replaced}
\zeta_{\rm ion}\to\zeta_{\rm ion}\left(1+\frac{\varepsilon_{\rm s}}{\varepsilon_{\rm heat}}\right)\,.
\end{equation}
Thus, Equations~(\ref{shift_s})--(\ref{zeta_replaced}) extend the results of the present paper by including additional
CR-induced mechanisms of dust heating.

Note that $T_{\rm g}$ in governing equations~(\ref{T_dust}) and (\ref{T_gas}) is to be replaced with its shifted value only
where it explicitly enters, i.e., Equation~(\ref{shift_s}) does not apply to the thermal velocity scale $v_{\rm g}^*(T_{\rm
g})$. Consequently, after substituting Equations~(\ref{shift_s})--(\ref{zeta_replaced}) into the analytical approximation,
Equations~(\ref{T_gas2}) and (\ref{Psi}), term $I_2\Delta T_{\rm g}$ in $\Psi(\Delta T_{\rm g})$ splits into two: the first
two components of $I_2$ in Equation~(\ref{I12}) multiplied with $\Delta T_{\rm g}$ plus the last component multiplied with
$(\Delta T_{\rm g}+T_{\rm g,s})$. Similarly, for the approach of effective grain radius, Equation~(\ref{Psi_eff}), term
$(\ldots+\ldots)\Delta T_{\rm g}$ inside the square brackets splits into the first component in the parentheses multiplied
with $\Delta T_{\rm g}$ plus the second component multiplied with $(\Delta T_{\rm g}+T_{\rm g,s})$.

The above analysis allows us to understand the relative importance of the surface and volume dust heating by CRs, depending
on the value of the critical grain radius $A$. From Equation~(\ref{Td_vs_Tg}) it follows that the effect of surface heating
is weak for grains much larger than $A$, i.e., their temperature is close to $(T_{\rm d0}^6+T_{\rm dv}^6)^{1/6}$ and the
contribution of $T_{\rm g,s}$ is negligible. Hence, for $A_0\ll a_{\rm eff}$ the gas temperature is determined from the
generalized standard model, Equation~(\ref{Psi_small}), with $T_{\rm d0}$ replaced according to Equation~(\ref{shift_v}). In
the opposite limit of $A_0\gg a_{\rm eff}$ one can rigorously show that $T_{\rm g,s}\lesssim(a_{\rm eff}/A_0)\Delta T_{\rm
g}$, i.e., $T_{\rm g,s}$ is negligible, too. The gas temperature in this case is described by Equation~(\ref{T_gas3}) with
$T_{\rm d0}$ and $\zeta_{\rm ion}$ replaced according to Equations~(\ref{shift_v}) and (\ref{zeta_replaced}). This shows
that the surface dust heating by CRs is only important for large $A_0/a_{\rm eff}$, where its effect is merely equivalent to
increasing the ionization rate.

\subsection{Role of the additional heating for dense cores}

In Section~\ref{example} we show that the generalized standard model reasonably describes $T_{\rm g}$ for conditions of the
pre-stellar core L1544, in the density range where the predicted difference between gas and dust temperatures should be
measurable. Therefore, for such cores the gas temperature practically does not depend on the surface heating; $T_{\rm g}$
could only be affected by the additional volume heating, which leads to higher $T_{\rm d0}$ according to
Equation~(\ref{shift_v}).

The absorption efficiency of the Lyman-Werner photons by silicate grains with $a\gtrsim0.01~\mu$m is approximately described
by a size-independent $Q_{\rm abs}$ \citep[][]{Draine2011Book}. Thus, UV radiation due to H$_2$ and He electronic excitation
by CRs mostly contributes to the surface heating, with $\varepsilon_{\rm s}\approx8$~eV per H$_2$ ionization
\citep[][]{Dalgarno1999,Glassgold2012}. (Formation of molecular hydrogen on grains adds to the surface heating, but the
resulting value of $\varepsilon_{\rm s}$ (of the order of a few eV) is quite uncertain.) Near-IR radiation from
vibrationally excited H$_2$ could potentially contribute to $\varepsilon_{\rm v}$ \citep[neglecting collisional quenching,
see][]{Dalgarno1999}, with up to $\approx3~$eV per H$_2$ ionization, while the effect of direct CR bombardment is negligible
\citep[unless $\zeta_{\rm ion}$ is extremely high, see][]{Hocuk2017}. By comparing Equation~(\ref{T_dv}) with
(\ref{T_deff}), noting that $\varepsilon_{\rm v}/\varepsilon_{\rm heat}<0.2$, and making use of parametrization~(\ref{rel2})
in Appendix~\ref{A3}, we conclude that the expected correction to $T_{\rm d0}^6$ cannot exceed a few percent.

\section{Appendix B\\ Functions $I_{1,2}$}
\label{A2}

Auxiliary functions $I_{1,2}(\tilde A,\tilde R, \gamma)$ entering Equation~(\ref{Psi}) are given by the following
expressions:
\begin{equation}\label{I12}
I_1=\int_1^{\tilde R}\frac{x^{-0.5+\gamma}}{x+\tilde A}\:dx\,,\quad
I_2=\frac12\left(I_1+\tilde A\frac{\partial I_1}{\partial \tilde A}+\frac52\tilde A^2\frac{\partial^2 I_1}{\partial \tilde A^2}\right)\,.
\end{equation}
For the MRN distribution ($\gamma=0$) we get
\begin{equation}\label{I1}
I_1=2\:\frac{\arctan\sqrt{\tilde A}-\arctan\sqrt{\tilde A/\tilde R }}{\sqrt{\tilde A}}\,.
\end{equation}
Generally, the integral in Equation~(\ref{I12}) can be calculated analytically for integer and half-integer $\gamma$.

\section{Appendix C\\ Parametrization for $T_{\rm g}$ and $T_{\rm d,eff}$}
\label{A3}

To calculate the gas temperature, we employ the general analytical approximation, Equations~(\ref{T_gas2}) and (\ref{Psi}),
with $I_1$ from Equation~(\ref{I1}). Figure~\ref{fig3} suggests that for typical conditions in dense cores the nonlinearity
in $T_{\rm g}$ is only significant for small values of $A_0/a_{\rm eff}$, asymptotically described by
Equation~(\ref{Psi_small}). Therefore, we use $v_{\rm g}^*(T_{\rm g})$ for the thermal velocity scale in
Equation~(\ref{T_gas2}) and neglect the term with $I_2$ in Equation~(\ref{Psi}); the latter introduces only a small error
(of about 6\% for $\zeta_{\rm ion}= 10^{-16}$~s$^{-1}$) in the regime of large $A_0/a_{\rm eff}$. After some manipulation,
we obtain
\begin{equation}\label{rel1}
\left(T_{\rm g}-T_{\rm d0}\right)\sqrt{T_{\rm g}}=\left(\frac{\zeta_{\rm ion}}{10^{-16}~{\rm s}^{-1}}\right)
\left(\frac{q}{\arctan q}\right)\frac{1+p_2}{p_1}\,,
\end{equation}
where $T_{\rm g}$ and $T_{\rm d0}$ are in units of Kelvin. Equation~(\ref{rel1}) depends on the following three parameters:
\begin{eqnarray*}
  p_1 &=& 0.0252\sqrt{\tilde R}\left(\frac{n_{\rm g}}{10^5~{\rm cm}^{-3}}\right), \\
  p_2 &=& 0.0115\sqrt{\tilde R}\left(\frac{n_{\rm g}}{10^5~{\rm cm}^{-3}}\right)\left(\frac{T_{\rm d0}}6\right)^{-5},\\
    q &=& \frac{\sqrt{p_2}}{1+p_2}\frac{\sqrt{\tilde R}-1}{\sqrt[4]{\tilde R}}\,,\\
\end{eqnarray*}
determined by $\tilde R=a_{\rm max}/a_{\rm min}$, where $a_{\rm max}=0.25~\mu$m is fixed. Parameter $q$ is a measure of the
size distribution width; the case of monodisperse dust is recovered in the limit $q\to0$, where $(q/\arctan q)\to1$. For
$p_2\ll1$, Equation~(\ref{rel1}) tends to the generalized standard model, Equation~(\ref{Psi_small}), for $p_2\gtrsim1$ it
approaches the universal asymptote of Equation~(\ref{T_gas3}).

The effective dust temperature is directly obtained from Equation~(\ref{T_deff}):
\begin{equation}\label{rel2}
T_{\rm d,eff}=T_{\rm d0}\left[1+0.202
\left(\frac{\zeta_{\rm ion}}{10^{-16}~{\rm s}^{-1}}\right)\left(\frac{T_{\rm d0}}6\right)^{-6}\right]^{1/6},
\end{equation}
where, again, $T_{\rm d,eff}$ and $T_{\rm d0}$ are in units of Kelvin.

\bibliographystyle{apj}
\bibliography{refs}

\end{document}